\documentclass[amsmath,preprintnumbers,noshowpacs,twocolumn,plb,superscriptaddress, nofootinbib]{revtex4}
\usepackage{amssymb}
\usepackage{graphicx}
\usepackage{epstopdf}
\usepackage{bm}
\usepackage{epsfig}
\usepackage{graphics}
\usepackage{xspace}
\usepackage{amsfonts}
\usepackage{verbatim}
\usepackage{natbib}
\usepackage{color}
\usepackage[none]{hyphenat}
\graphicspath{{./figures/}}

\begin{document}

\title{Nuclear matter effects on jet production at electron-ion colliders}

\author{Hai Tao Li}
\email{haitao.li@northwestern.edu}
\affiliation{HEP Division, Argonne National Laboratory, Argonne, Illinois 60439, USA}
\affiliation{Department of Physics \& Astronomy, Northwestern University, Evanston, Illinois 60208, USA }
\affiliation{Theoretical Division, Los Alamos National Laboratory, Los Alamos, NM, 87545, USA}
\author{Ivan Vitev}
\email{ivitev@lanl.gov}
\affiliation{Theoretical Division, Los Alamos National Laboratory, Los Alamos, NM, 87545, USA}

\begin{abstract}
Jet production and jet substructure in reactions with nuclei at future electron-ion colliders will play a preeminent role in the exploration of nuclear structure and the evolution of parton showers in strongly-interacting matter. In the framework of soft-collinear effective theory, generalized to include in-medium interactions, we present the first theoretical study of inclusive jet cross sections and the jet charge at the EIC. 
Predictions for the modification of these observables in electron-gold relative to electron-proton collisions reveal how the flexible center-of-mass energies and kinematic coverage at this new facility can be used to enhance the signal and maximize the impact of the electron-nucleus program.  Importantly, we demonstrate theoretically how to disentangle the effects from nuclear parton distribution functions and the ones that arise from strong final-state interactions between the jet and the nuclear medium.   
\end{abstract} 

\maketitle

{\it Introduction.---} 
In the past decade, jets have emerged as premier diagnostics of the  properties of hot nuclear matter  created in  heavy-ion collisions.
The predicted synergy between the jet cross section suppression and  jet substructure modification~\cite{Vitev:2008rz,Vitev:2009rd}  has been a cornerstone of an ever growing experimental program at the Relativistic Heavy Ion Collider and the Large Hadron Collider. Examples of such recent and planned jet measurements can be found in the following Refs.~\cite{Timilsina:2016sjv,Adare:2015kwa,Adamczyk:2017yhe,Kauder:2017cvz,Adam:2020wen,Acharya:2018uvf,Acharya:2019jyg,Aaboud:2018twu,Aaboud:2019oac,Sirunyan:2018ncy,Sirunyan:2020qvi}.  
At the future high-luminosity and high-energy electron-ion collider (EIC), jet production will also play a vital role -- in precision tests of QCD,  the next generation studies of nucleon and nuclear structure,  and cold nuclear matter tomography.  
Recently, various jet observables have been investigated in deep inelastic scattering  (DIS), many of them with an eye on the EIC~\cite{Abelof:2016pby,Boughezal:2018azh,Gehrmann:2018odt,Hoche:2018gti,Dumitru:2018kuw,Liu:2018trl,Aschenauer:2019uex,Kang:2020fka,Li:2020bub,Arratia:2020azl, Kolbe:2020tlq}. 
The overwhelming majority of those studies have focused on electron-proton (e+p) reactions. 
In electron-ion (e+A) reactions, jet observables at the EIC have been discussed in the context of small-$x$ gluon saturation physics and lepton-tagged jet acoplanarity.

To develop  the e+A  jet physics program at the EIC,  it is prudent to place emphasis on  observables that have been the most illuminating and impactful in the case of heavy-ion collisions.  Their  measurement at the EIC is  expected to avoid most of the heavy-ion background subtraction challenges  due to the much cleaner DIS environment.  Modification of jet observables in reactions with nuclei  relative to the e+p case may arise from initial-state effects and final-state jet-medium interactions.  In this pioneering study, our goal is to maximize and isolate the latter. 
Recent work on light and heavy meson cross section modification at the EIC~\cite{Li:2020zbk} has shown that it is advantageous to focus on the forward proton/nucleus going direction. The relevant EIC center-of-mass  $\sqrt{s}$ energies and pseudorapidity  $\eta$  correspond to moderate and large values of Bjorken-$x$ -- away from the gluon saturation regime. Still, initial-state effects are expected to be present as encoded by nuclear parton distribution functions (nPDFs), see Ref.~\cite{Ethier:2020way} for a recent review. It is crucial to investigate observables  which can disentangle these two distinct physics contributions.

In this work, we carry out the first calculation of inclusive jet production and the jet charge in electron-nucleus collisions at the EIC and investigate the impact of initial-state and final-state cold nuclear matter effects.  The inclusive jet cross section can be expressed in a  factorized form with the help of semi-inclusive jet functions (SiJFs), 
 \begin{align}
E_{J} \frac{d^{3} \sigma^{l N \rightarrow J X}}{d^{3} P_{J}} &=\frac{1}{S} \sum_{i, f} \int_{0}^{1} \frac{d x}{x} \int_{0}^{1} \frac{d z}{z^{2}} f_{i / N}(x, \mu) \nonumber \\
& \times  \hat{\sigma}^{i \rightarrow f}(s, t, u, \mu)J_{f }(z, p_T R, \mu) \;, 
\end{align}
where $ f_{i / N}$ is the PDF of parton $i$ in nucleon $N$.  $\hat{\sigma}^{i\to f}$ is the partonic cross section with initial state parton $i$  and final state parton $f$, which we take  up to NLO in QCD~\cite{Hinderer:2015hra}.  $J_{f}$ is the SiJF initialed by parton $f$,  it  can be found in Refs.~\cite{Kang:2016mcy,Dai:2018ywt} and was derived in the framework the soft-collinear effective theory (SCET)~\cite{Bauer:2000ew,Bauer:2001yt,Beneke:2002ph}. 
When the jet radius $R$ is small,  potentially large logarithms of the type $\ln R$ can be resummed  by  evolving  the  jet  function from the jet scale $p_TR$ to the factorization scale $\mu$.  Parton shower formation in nuclear matter has been described in the  framework of soft-collinear effective theory with Glauber gluon interaction (SCET$_{\rm G}$)~\cite{Ovanesyan:2011xy,Kang:2016ofv}. Hence, the SiJF formalism  can be extended  
 to heavy-ion collisions, as was first demonstrated in Refs.~\cite{Kang:2017frl,Li:2018xuv}  for the cross sections of light and heavy flavor jet production. This universal approach has been validated by experimental measurements, most recently on the jet radius dependence of jet cross 
 sections~\cite{CMS:2019btm},   and can be applied to e+A collisions.

Different from inclusive jet cross sections, jet substructure measures the radiation pattern inside a given jet and is governed by a smaller intrinsic scale.  
Even though the differences between the substructure of jets in e+p and e+A are expected to be smaller than the differences between p+p and  A+A collisions, we will show on the example of the jet charge~\cite{Field:1977fa}  that  nuclear effects can be  identified.  The average jet charge is defined as the transverse momentum $p_T^{i}$  weighted sum of the charges $Q_i $ of the jet constituents
\begin{align} \label{eq:charge}
    Q_{\kappa, {\rm jet}}  = \frac{1}{\left(p_T^{\rm jet}\right)^\kappa } \sum_{\rm i\in jet} Q_i \left(p_T^{i} \right)^{\kappa } \; ,  \quad \kappa > 0 \; ,
\end{align}
where $\kappa$ is a free parameter that must be positive for infrared safety.  Its value controls the relative contribution of soft and hard particles to   $Q_{\kappa, {\rm jet}}$ --  a larger $\kappa$ would suppress direct contribution from soft particles. 
Studies in proton and heavy-ion collisions~\cite{Krohn:2012fg,Li:2019dre,Sirunyan:2020qvi} have found that the jet charge is strongly correlated with the electric charge of the  parent parton and can be used to separate  quark jets from anti-quark jets and to pinpoint their flavor origin.
A key point of this theoretical work  is to demonstrate how to disentangle initial-state effects and final-state effects for the  inclusive jet cross section and the jet charge. 

{\it Cold Nuclear Matter Effects.---} 
In this work initial-state effects are  included through global-fit nuclear PDFs~\cite{Kovarik:2015cma} that parametrize the inclusive DIS cross section modification in the  EMC, anti-shadowing, and shadowing regions into leading-twist distributions with the assumption of collinear factorization.  Isospin symmetry is implemented on account of the fact that the nucleus is a mix of protons and neutrons, $A=Z+N$,  which will change the total density for up and down quarks significantly in the nuclear PDFs  relative to the proton ones.  Clearly, it will be important to investigate a possible way for efficient flavor tagging  to test isospin symmetry in large nuclei and get a better handle on the flavor dependence of nuclear effects.

To account for  final-state interactions, we  make use of the medium induced splitting kernels  derived in the framework of SCET$_{\rm G}$~\cite{Ovanesyan:2011xy,Ovanesyan:2011kn} and verified using a lightcone wavefunction approach with DIS applications in mind~\cite{Sievert:2018imd,Sievert:2019cwq}.  These splitting kernels capture the medium effects on the full  collinear shower dynamics. 
Similar to the squared matrix elements  for collinear splitting, the real contribution for  the medium corrections to the $i\to jk$ branching process  with  identified initial state parton $i$ and final state parton $j$ we denote as 
$   f_{i \rightarrow j k }^{\mathrm{med}} \left(z, \mathbf{k}_{\perp}\right) = {dN_{i \rightarrow j k }^{\mathrm{med}} }/{d^2\mathbf{k}_{\perp}  dz}.$ The virtual contributions are obtained following Ref.~\cite{Collins:1988wj}. The collinear splitting kernels are singular when $z\to1$, if $i=j$.  It is important to note that 
\begin{equation} 
 f_{i \rightarrow j k }^{\mathrm{tot}} \left(z, \mathbf{k}_{\perp}\right)  =  f_{i \rightarrow j k }^{\mathrm{vac}} \left(z, \mathbf{k}_{\perp}\right)  +  f_{i \rightarrow j k }^{\mathrm{med}} \left(z, \mathbf{k}_{\perp}\right) \;, 
\end{equation}
which in turn leads to a medium-induced contribution to the SiJFs and the evolution of fragmentation functions.

Because in-medium splitting kernels are calculated by integrating over the interactions in matter and the jet propagation path, they can only be  obtained numerically and analytic dimensional regularization cannot be applied.  
The medium corrections to the SiJFs are obtained with an ultraviolet cut off at the scale $\mu$ corresponding to the factorization scale  as implemented  in Refs.~\cite{Kang:2017frl, Li:2018xuv}. In analogy to SiJFs in vacuum, we take the relevant real and virtual contributions inside and outside of the jet cone to find
\begin{align}\label{eq:sp}
   J_{q}^{\rm{med}}\left(z, p_T R, \mu \right)&=\left[\int_{z(1-z) p_T R}^{\mu} d^2\mathbf{k}_{\perp} f_{q \rightarrow q g}^{\mathrm{med}} \left(z, \mathbf{k}_{\perp}\right)\right]_{+}
    \nonumber \\ 
    &+\int_{z(1-z) p_T R}^{\mu} d^2\mathbf{k}_{\perp} f_{q \rightarrow gq}^{\mathrm{med}} \left(z, \mathbf{k}_{\perp}\right)  \; ,
\end{align}
and we use $2 E_J \tan R/2\cosh \eta \approx p_T R$ in Eq.~(\ref{eq:sp}). 
A similar expression can be derived for the gluon SiJF, but we note that in the high $p_T$ and forward rapidity EIC kinematics that we are interested in gluon contribution to jet production is insignificant even at NLO. 
In SiJFs all  singularities when $z\to 1$ are regularized by the plus-distribution function that has the standard definition 
  $  \int_0^1 dz\, g(z)\left[f(z)\right]_+ =\int_0^1 dz \left(g(z)-g(1)\right) f(z)\;. $

Moving on  to the jet charge modification at the EIC,  to the order that we work the average gluon jet charge is zero due to electric charge conservation. The   quark jet charge can be derived in SCET from the collinear factorization formula for measuring a hadron inside a jet~\cite{Krohn:2012fg}
\begin{align} \label{eq:Q}
     \langle Q_{\kappa,q} \rangle =& \frac{\tilde{\mathcal{J}}_{qq}(E,R,\kappa,\mu)}{J_{q}(E,R,\mu)} \tilde{D}_q^{Q}(\kappa) 
    \nonumber \\
     &  \times \exp\left[\int_{1{\rm GeV}}^{\mu}\frac{d{\mu'}}{{\mu'}}\frac{\alpha_s({\mu'})}{\pi}  \tilde{f}^{\rm vac}_{q \rightarrow q g}(\kappa) \right] \;, 
\end{align}
where $J_{q}(E,R,\mu)$ is a jet function and $\tilde{\mathcal{J}}_{qq}(E,R,\kappa,\mu)$ is the $(\kappa +1)$-th  Mellin moment of the Wilson coefficient for  matching  the quark  fragmenting  jet  function  onto a quark fragmentation  function.
The perturbative NLO jet function and the matching coefficients from the jet to the hadron can be found in Refs.~\cite{Ellis:2010rwa,Jain:2011xz}. Given $\kappa$, for each jet flavor the average jet charge only depends on one non-perturbative parameter $\tilde{D}_q^{Q}(\kappa)$, which is obtained from PYTHIA~\cite{Sjostrand:2014zea} simulations,  and the initial scale for the vacuum fragmentation function is set to 1~GeV.  In the above equation  $\tilde{f}^{\rm vac}_{q \rightarrow q g}(\kappa)$  is the $(\kappa+1)$-th Mellin moment of the splitting function ${f}^{\rm vac}_{q \rightarrow q g}(z)$. We found very good agreement between Eq.~(\ref{eq:Q}) and PYTHIA simulation for a large range of jet $p_T$ in both e+p and p+p collisions. 
We use the same non-perturbative parameter $\tilde{D}_q^{Q}(\kappa)$ for different jet radii $R$.

Nuclear matter effects on the jet charge were studied in Refs.~\cite{Chen:2019gqo,Li:2019dre} for the case of heavy-ion collisions. Following   the derivations in Ref.~\cite{Li:2019dre} the average jet charge at the EIC can be written as
\begin{align} \label{eq:AAQ}
    \langle Q_{\kappa, q}^{\rm eA} \rangle = &\langle Q_{\kappa,q}^{\rm ep} \rangle  \exp\left[ \int_{\mu_0}^{\mu} \frac{d{\mu'}}{{\mu'}}  
      \frac{\alpha_s({\mu'})}{2\pi^2} (2 \pi {\mu'}^2) \tilde{f}^{\rm med}_{q \rightarrow qg}(\kappa,{\mu'}) \right] 
    \nonumber \\ & \times  \left(1+\tilde{\mathcal{J}}^{\rm med}_{qq}-J_{q}^{\rm med}\right) 
   + \mathcal{O}(\alpha_s^2)\,.
\end{align}
Here,  the exponential term comes from the medium-modified DGLAP evolution from $\mu_0\approx \Lambda_{\rm QCD}$ to the jet scale 
and $\tilde{f}^{\rm med}_{q\rightarrow qg}(\kappa,\mu)=\int_0^1\,dx\, (x^\kappa-1) \, f^{\rm med}_{q \rightarrow qg}(x,\mu)$. 
Finally, from the second line of Eq.~(\ref{eq:AAQ}) we have explicitly
\begin{multline}
    \tilde{\mathcal{J}}^{\rm med}_{qq}-J_{q}^{\rm med} = \frac{\alpha_s(\mu)}{\pi} \int_0^1 dx~(x^\kappa-1)
    \\ \times
    \int_{0}^{2 E x(1-x)\tan R/2} d^2 \mathbf{k}_{\perp} 
    f_{q \rightarrow  qg }^{\rm{med}}\left(x, \mathbf{k}_{\perp} \right)\,.
\end{multline}

{\it Numerical Results.---}
In the calculations that follow we  use  the CT14nlo PDF sets~\cite{Dulat:2015mca} for the proton and the nCTEQ15FullNuc PDF sets~\cite{Kovarik:2015cma} for  the nucleus, as  provided by  {\sc \small Lhapdf6}~\cite{Buckley:2014ana}.  Consistent with  Ref.~\cite{Li:2020zbk},  we fix the  nominal transport coefficient of cold nuclear matter $\langle k_\perp^2\rangle/\lambda_g=0.12$ GeV$^2$/fm, consider a gold (Au) nucleus, and average over the nuclear geometry. The in-medium shower corrections induced by the interactions between the final-state parton and the nucleus vary with the parton energy in the nuclear rest frame, where the lower energy partons receive larger medium corrections. Therefore, we focus on  jet production in the forward rapidity region $2<\eta<4$, where the measurement is still possible but  the jet energy is lower in the nuclear rest frame.  For the inclusive jet cross section, we include  all  partonic channels  and the resolved photon contribution. Our results in e+p collisions are consistent with the ones from Ref.~\cite{Boughezal:2018azh}. 

\begin{figure}[t!]
	\centering
	\includegraphics[width=0.52\textwidth]{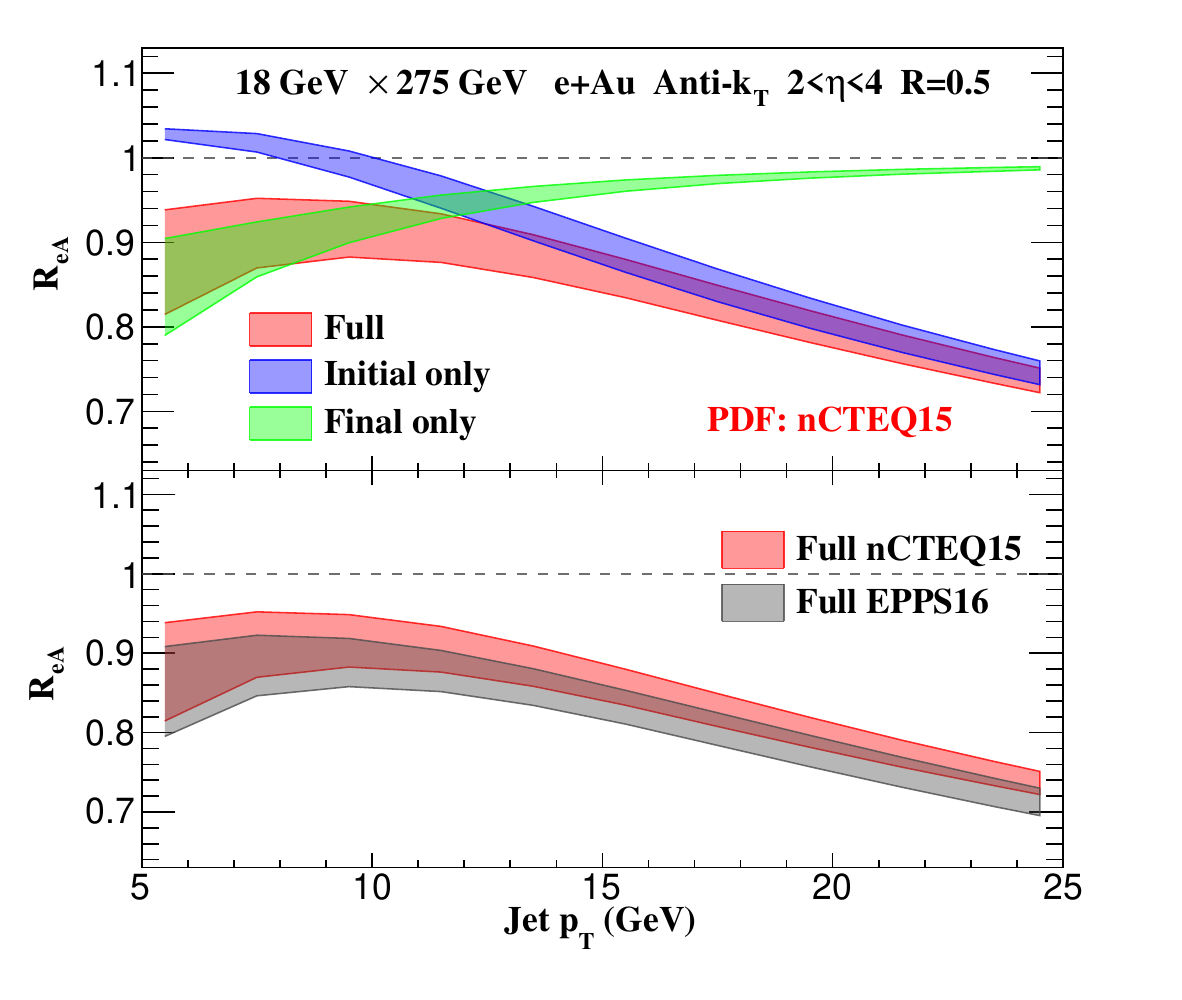}
	\vspace{-1cm}
	\caption{Modifications of the inclusive jet cross section in  18 $\times$ 275 GeV  e+Au collisions for the rapidity interval $2< \eta < 4$. 
 In the upper panel, the blue and green bands represent contributions from initial-state PDFs and final-state interaction between the jet and cold nuclear matter, while the red band is the full result. The lower panel shows the full $R_{eA}$  for two different nPDF sets. }
	\label{fig:ReAIF}
\end{figure}

Nuclear effects on  reconstructed jets in electron-nucleus collisions can be studied through the ratio 
\begin{align}
    R_{\rm eA}(R) = \frac{1}{A}   \frac{\int_{\eta1}^{\eta2} d\sigma/d\eta dp_T\bigskip
    |_{e+A}}{   \int_{\eta1}^{\eta2}  d\sigma/d\eta dp_T\big|_{e+p}}\,.
\end{align}
The jet calculations correspond to the  anti-k$_T$ algorithm and as a first example we choose a radius parameter $R=$0.5. The uncertainties of $R_{\rm eA}$ are calculated by varying the scale settings in the numerator and denominator simultaneously,  i.e. in a correlated way as it minimizes the variation due to the overall normalization of cross sections. In Fig.~\ref{fig:ReAIF} bands correspond to scale uncertainties from varying the factorization scale and the jet scale by a factor of two independently.  
For jet rapidity $\eta=2$ at leading order,  when the jet transverse momentum is in the range [5,25] GeV, the Bjorken-$x$ varies from  [0.09, 0.43] corresponding to the so-called anti-shadowing and EMC regions of nuclear PDFs. As a result, there is an enhancement for small $p_T$ due to anti-shadowing  and a suppression for large $p_T$ due to  the EMC effect, which is  shown by the blue band in the upper panel of Fig.~\ref{fig:ReAIF}.  The green band represents the final-state effects, which give rise to 10 - 20\% suppression when $p_T\sim 5$ GeV.  They are smaller for larger jet energy as expected, and going to backward rapidities further reduces the effect of medium-induced parton showers.  The predicted  full   $R_{\rm eA}(R=0.5)$ for 18 GeV (e) $\times$ 275 GeV (A) collisions  is given by the red band.  To illustrate the impact of a different nPDF choice, we show in the lower panel of  Fig.~\ref{fig:ReAIF}  a comparison between the $R_{eA}$ computed with the nCTEQ15~\cite{Kovarik:2015cma} and EPPS16~\cite{Eskola:2016oht} sets. We find that the difference in cross sections is less than 5\% \footnote{Other uncertainties can arise from Monte Carlo replicas within the same PDF set or variation in the transport properties of nuclear matter.}.  The measurements of  jet modification in the future will improve our understanding of strong interactions inside nuclei and nuclear PDFs at moderate and large Bjorken-$x$. 

 \begin{figure}[!t]
    \centering
    \includegraphics[width=0.5\textwidth]{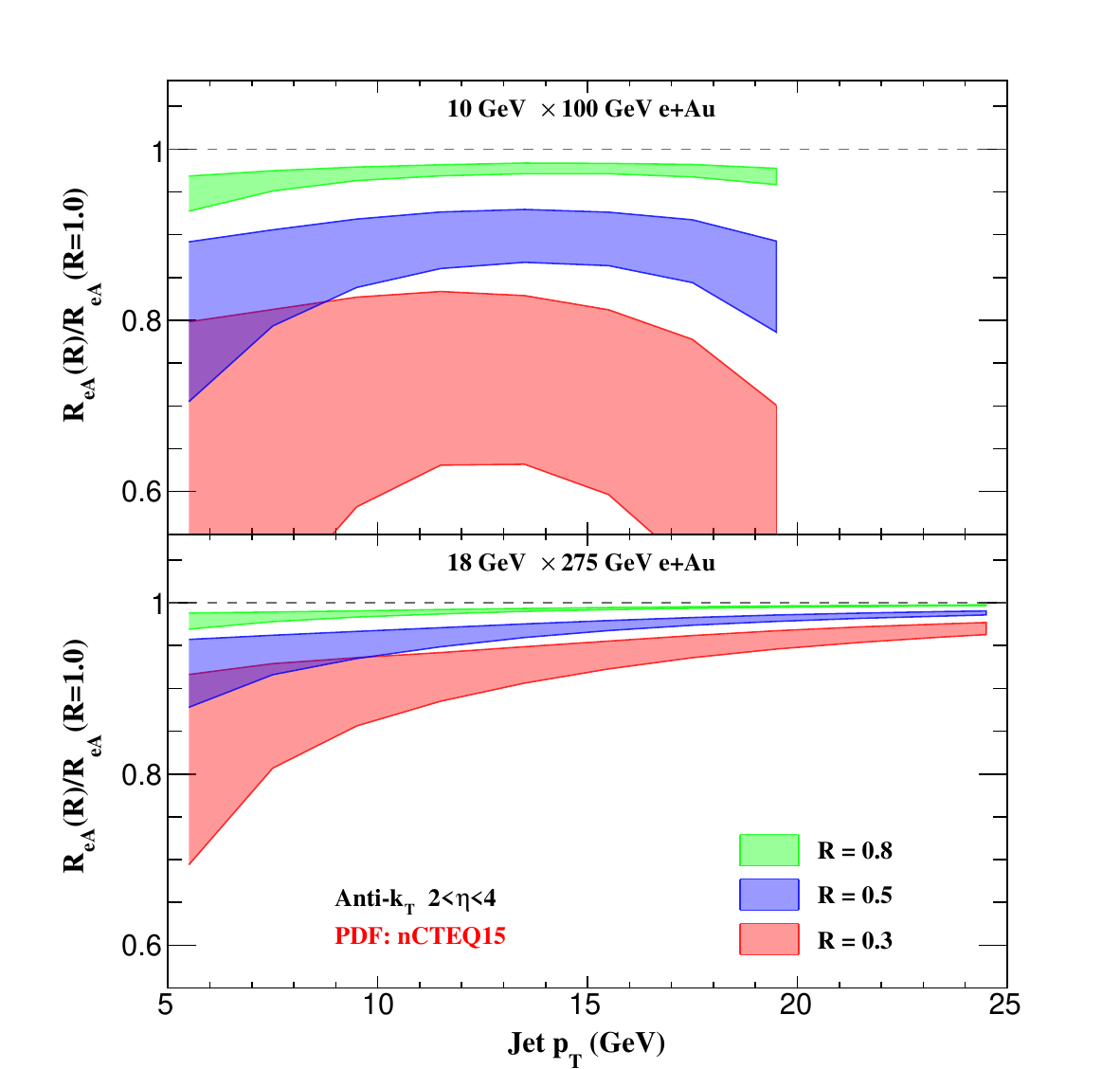}
    \vspace{-0.7cm}
    \caption{Ratio of jet  cross section modifications for different  radii $R_{\rm eA}(R)/R_{\rm eA}(R=1.0)$ in 10 $\times$ 100 GeV (upper) and  18 $\times$ 275 GeV (lower) e+Au collisions, where the smaller jet radius is R=0.3, 0.5, and 0.8,  and the jet rapidity  interval is $2<\eta<4$.   }
    \label{fig:ReAto1}
\end{figure}

To study cold nuclear matter transport properties with jets at the EIC, it is essential to reduce the role of nPDFs and enhance the effects due to final-state interactions.  An efficient strategy  is to measure the ratio of the modifications with different jet radii,  $R_{\rm eA}(R)/R_{\rm eA}(R=1)$, as for jets with the same kinematics initial-state effects in e+A reactions will cancel. This is also an observable  very sensitive to the details of  in-medium branching processes~\cite{Vitev:2008rz} and greatly discriminating with respect to theoretical models~\cite{CMS:2019btm}.   Furthermore, it is  very beneficial  to explore smaller center-of-mass energies  where the final-state effects are expected to be larger even though the cross section is  smaller. Such measurements will take advantage of the high-luminosity design of the future facility.  Our  predictions for the ratio of jet cross section suppressions for different radii at the EIC  is presented in Fig.~\ref{fig:ReAto1},  where the upper and lower panels correspond to  results for 10 GeV (e) $\times$ 100 GeV (A) and 18 GeV (e) $\times$ 275 GeV (A) collisions, respectively.  The plot in the upper panel is truncated around $p_T \sim 20$ GeV because of phase space  constraints  in the lower energy  collisions.    

By comparing the 18 GeV $\times$ 275 GeV e+Au collision results to the ones in Fig.~\ref{fig:ReAIF}  we see that $R_{\rm eA}(R)/R_{\rm eA}(R=1)$  indeed eliminates 
initial-state effects.  To underscore this point, in addition to using  the nCTEQ15 nPDF set~\cite{Kovarik:2015cma},  we evaluated the double ratio with the EPPS16~\cite{Eskola:2016oht} parameterization and found that the results  are indistinguishable.  The red, blue, and green bands denote ratios with $R=0.3\,,0.5\,,0.8$, respectively. 
Since medium-induced parton showers are broader than the ones in the vacuum,   for smaller jet radii the suppression from final-state interactions is more significant.   
Even though the scale uncertainties also grow,  the nuclear effect is very clear and its magnitude is further significantly enhanced  by the steeper $p_T$ spectra at lower $\sqrt{s}$.

\begin{figure}
    \centering
    \includegraphics[width=0.5\textwidth]{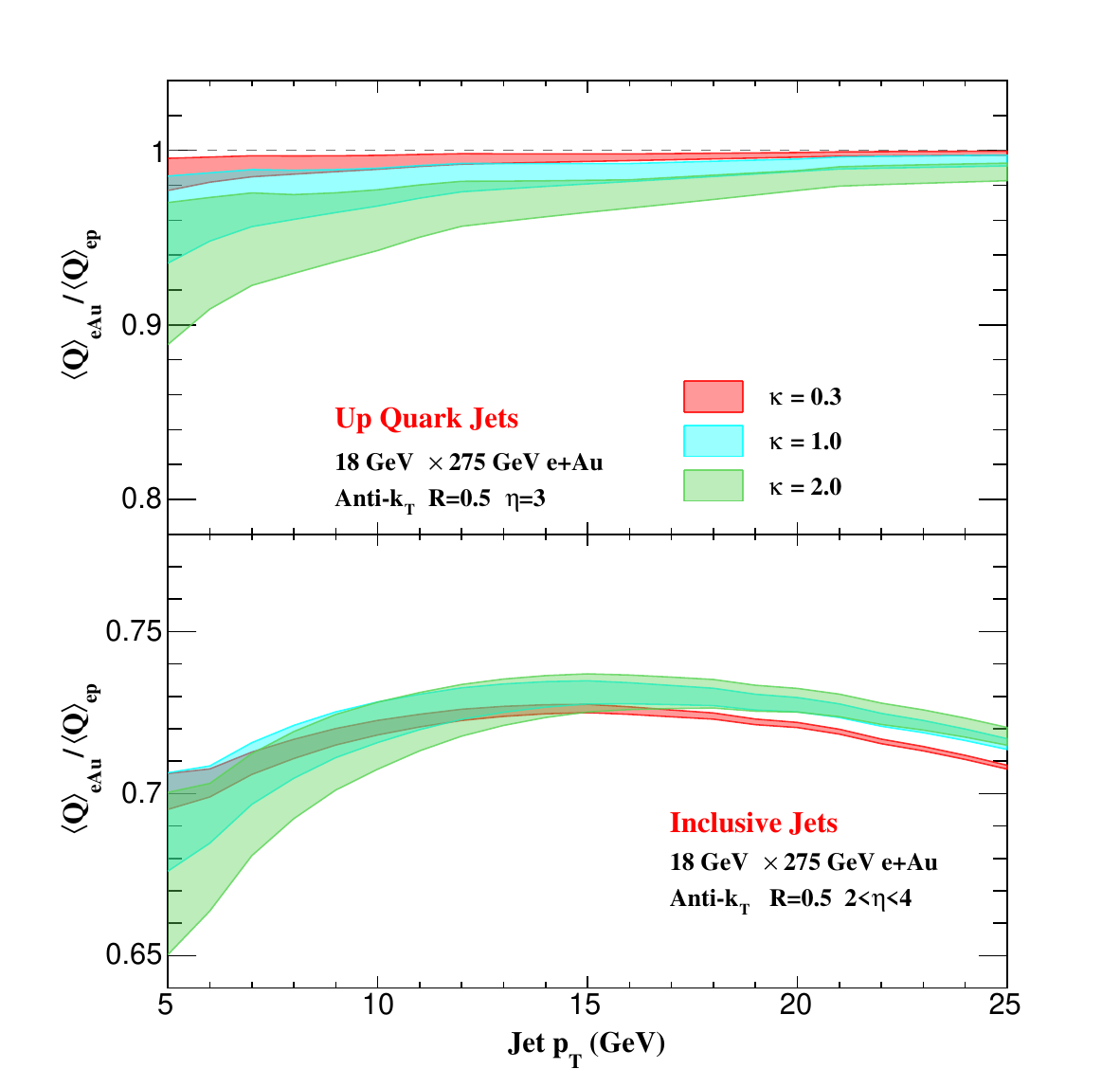}
    \caption{Modifications of the jet charge in e+Au collisions. The upper panel is the modification for up-quark jet with $\eta=3$ 
    in the lab frame. The lower panel is the results for inclusive jet with $2<\eta<4$ in 18 $\times$ 275 GeV e+Au collisions.  }
    \label{fig:charge}
\end{figure}

For jet substructure, Fig.~\ref{fig:charge} presents our jet charge results at the EIC   in 18 GeV $\times$ 275 GeV e+Au collision and for radius parameter $R=0.5$.  
The red, blue and green bands correspond to the jet charge parameter $\kappa=0.3\,,1.0\,,2.0$, see Eq.~(\ref{eq:charge}),   respectively.  
The upper panel shows the modification for the average charge of up-quark initialed jets, where the rapidity is fixed to be  $\eta=3$. It  is defined as  $\langle Q_{ \kappa,q}^{\rm eA} \rangle/\langle Q_{\kappa,q}^{\rm ep} \rangle$ and predicted by Eq.~(\ref{eq:AAQ}),  which is independent of the jet flavor and 
originates purely  from final-state interactions.  Flavor separation for jets has  been accomplished at the LHC~\cite{Aad:2015cua} and  should be pursued at the EIC.
For a larger $\kappa$, the $(\kappa+1)$-th Mellin moment of the splitting function is more sensitive to  soft-gluon emission in that it affects the  $z\sim 1$ region in the splitting function where medium enhancement for soft-gluon radiation is the largest.  As shown in the upper panel of Fig.~\ref{fig:charge}, the modification is more significant for larger $\kappa$. The overall corrections are of order 10\% or smaller and  decrease with increasing $p_T$. 
The modification of the average charge for inclusive jets behaves very differently because there is a  cancellation between contributions from jets initiated by  different flavor partons, in particular from up quarks and down quarks. The lower panel of Fig.~\ref{fig:charge} shows the ratio of average charges for inclusive jets with $R=0.5$ and $2<\eta<4$ for e+A and e+p collisions.  The modification is about 30\% and the $\kappa$ dependence is small due to the large difference between up/down quark density between proton and gold PDFs. Precision measurement of the charge for inclusive jets will be an excellent way to constrain isospin effects and  the up/down quark PDFs in the nucleus.

{\it Conclusions.---}  
In summary, we presented a pioneering  study of inclusive jet cross sections and the jet charge in electron-nucleus collisions at the EIC building upon the SCET approach. Initial-state effects were considered via global-fit nuclear PDFs and the corrections that arise from interactions between the jet and nuclear matter were implemented with the help of the medium-induced splitting kernels derived in the framework of SCET with Glauber gluon interactions. Our results demonstrate that the forward proton/nucleus going direction is the optimal region to observe large final-state modifications due to in-medium shower evolution. We further find that final-state effects on $R_{\rm eA}$ are large in the relatively small  $p_T$ region, whereas initial-state effects, if sizeable, are observed at large $p_T$.  A major advantage of jet measurements in comparison to the ones for semi-inclusive hadron production is that by considering the ratio of cross section modifications for different jet radii  the effects from nuclear PDFs can be strongly suppressed to cleanly  probe the strong interaction between jets and cold nuclear matter.  With judicious choice of  the center-of-mass energy, rapidity interval, and jet radius  $R$, the inclusive cross section suppression can be nearly a factor of two -- similar to what is measured with high precision in A+A relative to p+p collisions. Related to the jet attenuation in cold nuclear matter is the modification of jet substructure. The jet charge modification of individual flavor jets can shed light on the medium-induced scaling violations in QCD,  whereas a precision study of the  charge of inclusive jets can be used to  extract the flavor information and constrain the nuclear PDFs. 

Our work is an essential step in defining the jet physics program in e+A collisions at the EIC~\cite{AbdulKhalek:2021gbh} and in guiding experimental focus at other proposed DIS facilities, such as the Large Hadron electron Collider (LHeC)~\cite{Klein:2018rhq}  and an Electron ion collider in China (EicC)~\cite{Chen:2018wyz}.  Results from this study suggest that the center-of-mass energies of order TeV at the LHeC  will eliminate medium-induced parton-shower effects  and the facility will be best suited to study nuclear PDFs and small-x physics. Conversely, the low center-of-mass energies at the EicC are very well suited to study final-state interactions in cold nuclear matter, though the $p_T$ of measurements will be limited.  The EIC occupies a sweet spot that ensures the broadest impact of its electron-nucleus program.

\begin{acknowledgments}
We would like to thank Z.L. Liu for useful discussions. H.T. Li and  I. Vitev are supported by the  LDRD program at LANL. 
I. Vitev is partly supported by the U.S. Department of Energy under Contract No. 89233218CNA000001. 
H.T. Li is partly supported  by   the U.S. Department of Energy under Contract  No. DE-AC02-06CH11357 and the National Science Foundation under Grant No. NSF-1740142.
 \end{acknowledgments}


\end{document}